\documentclass[aps,prl,superscriptaddress,twocolumn,showpacs,preprintnumbers,amsmath,amssymb]{revtex4}

\usepackage{graphicx}
\usepackage{longtable}
\usepackage{bm}% bold math
\usepackage{multirow}
\usepackage{hyperref}
\usepackage{chngcntr}
\begin{document}

\title{Nematic Fluctuations and Phase Transitions in LaFeAsO: a Raman Scattering Study}

\author{U. F. Kaneko}
\affiliation{``Gleb Wataghin'' Institute of Physics, University of Campinas - UNICAMP, Campinas, S\~ao Paulo 13083-859, Brazil}

\author{P. F. Gomes}
\affiliation{``Gleb Wataghin'' Institute of Physics, University of Campinas - UNICAMP, Campinas, S\~ao Paulo 13083-859, Brazil}
\affiliation{Federal University of Goi\'as - UFG, Jata\'i, Goi\'as 75801-615, Brazil}

\author{A. F. Garc\'ia-Flores}
\affiliation{Federal University of ABC - UFABC, Santo Andr\'e, S\~ao Paulo 09210-580, Brazil}

\author{J.-Q. Yan\footnote{Present address: Materials Science and Technology Division, Oak Ridge National Laboratory, Oak Ridge, Tennessee 37831, USA.}}
\affiliation{Ames Laboratory, US-DOE, Ames, Iowa 50011, USA}

\author{T. A. Lograsso}
\affiliation{Ames Laboratory, US-DOE, Ames, Iowa 50011, USA}
\affiliation{Department of Materials Sciences and Engineering, Iowa State University, Ames, Iowa 50011, USA.}

\author{G. E. Barberis}
\affiliation{``Gleb Wataghin'' Institute of Physics, University of Campinas - UNICAMP, Campinas, S\~ao Paulo 13083-859, Brazil}

\author{D. Vaknin}
\affiliation{Ames Laboratory, US-DOE, Ames, Iowa 50011, USA}
\affiliation{Department of Physics and Astronomy, Iowa State University, Ames, Iowa 50011, USA}

\author{E. Granado}
\affiliation{``Gleb Wataghin'' Institute of Physics, University of Campinas - UNICAMP, Campinas, S\~ao Paulo 13083-859, Brazil}

\begin{abstract}
	
	Raman scattering experiments on LaFeAsO with splitted antiferromagnetic ($T_{AFM}=140$ K) and tetragonal-orthorhombic ($T_S=155$ K) transitions show a quasi-elastic peak (QEP) in $B_{2g}$ symmetry (2 Fe tetragonal cell) that fades away below $\sim T_{AFM}$ and is ascribed to electronic nematic fluctuations. A scaling of the reported shear modulus with the $T-$dependence of the QEP height rather than the QEP area indicates that magnetic degrees of freedom drive the structural transition. The large separation between $T_S$ and $T_{AFM}$ in LaFeAsO compared with their coincidence in BaFe$_2$As$_2$ manifests itself in slower dynamics of nematic fluctuations in the former.
	
\end{abstract}

\maketitle

\begin{figure}
	\includegraphics[width=0.4 \textwidth]{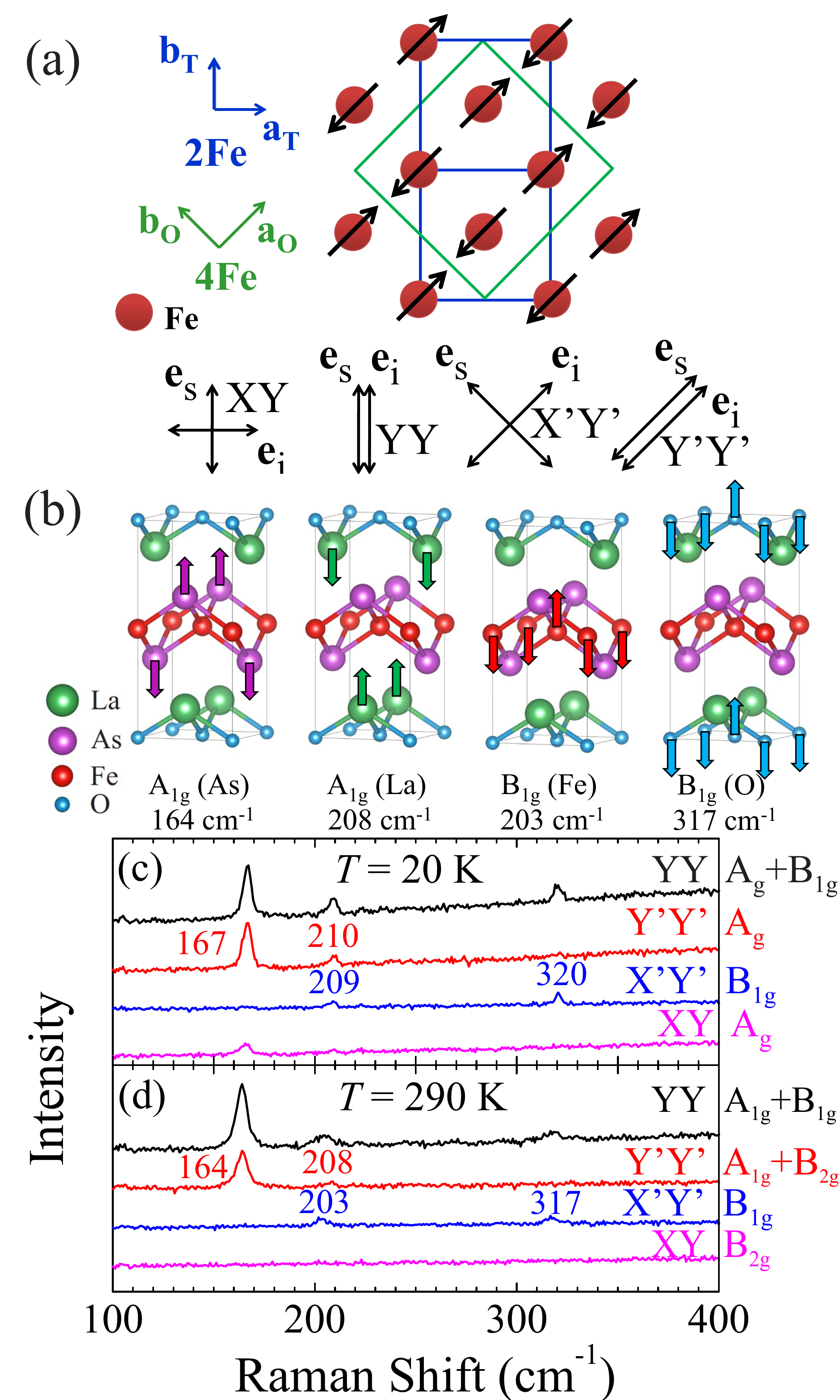}
	\caption{\label{Phonons} (Color online) (a) Schematic view of the Fe square lattice with the stripe antiferromagnetic structure and representations of the $XY$, $YY$, $X'Y'$ and $Y'Y'$ linear polarizations. The unit vectors {\bf e}$_i$ and {\bf e}$_s$ represent the polarizations of the incident and scattered photons, respectively. The edges and axes for the 2 Fe tetragonal and 4 Fe orthorhombic unit cells are  also displayed; (b) Raman-active phonons accessible in the scattering geometry employed in this work. The corresponding symmetries and observed frequencies at 290 K are indicated; (c,d) Raman spectra for distinct polarizations at $T=20$ K (c) and $T=290$ K (d). In (c) and (d), the symmetry associated with each polarization is given with respect to the corresponding orthorhombic and tetragonal unit cells, and the employed laser power was 10 mW.}
\end{figure}

The discovery of Fe-based superconductors (FeSCs) with high transition temperatures (above 100 K in FeSe films \cite{Ge}) triggered much interest on these materials \cite{Kamihara,Paglione,Fernandesnem,Fernandes}. Nematicity, characterized by large in-plane electronic transport anisotropy \cite{Chu}, is normally observed below a tetragonal-orthorhombic transition temperature $T_S$, and seems to be also present in other high-$T_c$ superconductors \cite{Ando}. Also, divergent nematic susceptibility in the optimal doping regime suggests that nematic fluctuations play an important role in the superconducting pairing mechanism \cite{Kuo}. Thus, investigations of the nematic order and fluctuations in FeSCs and their parent materials are pivotal to unraveling the origin of high-$T_c$ superconductivity. Clearly, it is necessary to identify the primary order parameter associated with the nematic phase \cite{Fernandesnem,Fernandes}. A relation between nematicity and magnetism is suggested by the near coincidence between $T_S$ and the antiferromagnetic (AFM) ordering temperature $T_{AFM}$ in some materials, most notably BaFe$_2$As$_2$ with $T_{AFM} \sim T_S = 138$ K \cite{Kim,Garitezi}. In fact, the magnetic ground state is a stripe AFM phase that breaks the 4-fold tetragonal symmetry of the lattice  (see Fig. \ref{Phonons}(a)), providing a natural mechanism for electronic anisotropy. On the other hand, $T_S$ and $T_{AFM}$ are significantly separated for LaFeAsO (LFAO) ($T_{AFM}=140$ K and $T_S=155$ K) \cite{McGuire,Yan2009,Li}, while FeSe does not order magnetically at ambient pressure but still shows a nematic transition at $T_S=90$ K \cite{McQueen}, motivating suggestions that the nematic transition may be driven by charge/orbital degrees of freedom rather than magnetism in the latter \cite{Baek,Bohmer}. However, even for FeSe the magnetic scenario may still apply \cite{Chubukov}. In Ba(Fe$_{1-x}$Co$_x$)$_2$As$_2$ and other doped systems, the splitting between $T_{AFM}$ and $T_S$ increases with doping \cite{Kim,Pratt}. Overall, the primary order parameter that drives the structural/nematic transition at $T_S$ and the dominating mechanism of $T_{AFM}/T_S$ separation in parent FeSCs are not fully settled yet.

Raman scattering was recently employed as a probe of nematic fluctuations in FeSCs and their parent materials. In $A$(Fe$_{1-x}$Co$_x$)$_2$As$_2$ ($A=$ Ca, Sr, Ba, Eu) \cite{Chauviere,Chauviere3,Gallais,Gallais2,Kretzschmar,Zhang2016,Bohm}, Ba$_{1-p}$K$_p$Fe$_2$As$_2$ \cite{Bohm}, FeSe \cite{Massat,Gnezdilov} and NaFe$_{1-x}$Co$_x$As \cite{Thorsmolle}, a quasi-elastic peak (QEP) with $B_{2g}$ symmetry (considering the 2 Fe tetragonal cell, see Fig. \ref{Phonons}(a)) has been observed and interpreted in terms of either charge/orbital \cite{Gallais,Gallais2,Zhang2016,Massat,Yamase} or spin \cite{Kretzschmar,Bohm,Khodas,Yamase2} nematic fluctuations. An unambiguous experimental identification of the nature of the fluctuations generating the $B_{2g}$ Raman QEP (charge/orbital or magnetic) is challenging due to the inherent coupling between the corresponding degrees of freedom. Despite such extensive investigations in several materials, no Raman study of the nematic fluctuations in the key parent compound LFAO has been carried out yet. In this work, we fill this gap and investigate in detail the temperature dependence of both electronic and phonon Raman scattering in LFAO.

Details of the synthesis procedure and basic characterization of the crystal employed in this work, showing $T_S = 155$ K and $T_{AFM} = 140$ K, are described elsewhere \cite{Yan2009,Zhang}. A fresh $ab$ surface with $\sim 1 \times 1$ mm$^2$ was obtained by cleaving the crystal and immediately mounting it at the cold finger of a closed-cycle He cryostat. The polarized Raman spectra were taken in quasi-backscattering geometry using the 488.0 nm line as exciting source focused into the $ab$ surface with a spot of $\sim 50$ $\mu$m diameter. A triple 1800 mm$^{-1}$ grating spectrometer equipped with a $L$N$_2$-cooled multichannel CCD detector was employed. The instrumental linewidth was $\sim 4$ cm$^{-1}$. Figure \ref{Phonons}(a) illustrates a square lattice of the Fe atoms and sets the conventions for polarizations. The 2 Fe tetragonal (space group P$4/mmm$) and 4 Fe orthorhombic (space group C$mma$) unit cells and axes in the $ab$ plane are also represented.

Symmetry analysis indicates that four Raman-active phonons are accessible by our experimental geometry in both tetragonal ($2A_{1g}$ and $2B_{1g}$) and orthorhombic ($2A_{g}$ and $2B_{1g}$) phases. Illustrations of such modes are given in Fig. \ref{Phonons}(b) (see also Ref. \cite{Hadjiev}). The raw Raman spectra in the phonon region at distinct linear polarizations are given in Figs. \ref{Phonons}(c) ($T=20$ K) and \ref{Phonons}(d) ($T=290$ K). The $B_{1g}$ modes observed at 203 and 317 cm$^{-1}$ at $T=290$ K are ascribed to Fe and O vibrations along $c$, respectively \cite{Hadjiev}, while the $A_{1g}$ modes at 164 and 208 cm$^{-1}$ are ascribed to As and La vibrations along $c$. The position of the 164 cm$^{-1}$ mode is comparable to that reported for the As mode in NaFeAs (163 cm$^{-1}$ \cite{Um}) and in $A$Fe$_2$As$_2$ ($A=$ Ca, Sr, Ba) (180-190 cm$^{-1}$ \cite{Chauviere3,Choi,Litvinchuk,Chauviere2}). The $T$-dependence of this phonon was investigated in detail (see SM). Its linewidth at low-$T$ is resolution-limited, suggesting a high crystalline quality, and shows a maximum at $T \sim T_{AFM}$ with no anomaly at $T_S$. Frequency anomalies are observed for this mode at both $T_S$ and $T_{AFM}$. Finally, an enhancement in $XY$ polarization is observed below $T_{AFM}$, which is similar to related systems \cite{Chauviere3,Choi,Chauviere2,Kretzschmar} and is due to the coupling of this phonon with anisotropic electronic states in the magnetic phase \cite{Martinez}.

\begin{figure}
	\includegraphics[width=0.5 \textwidth]{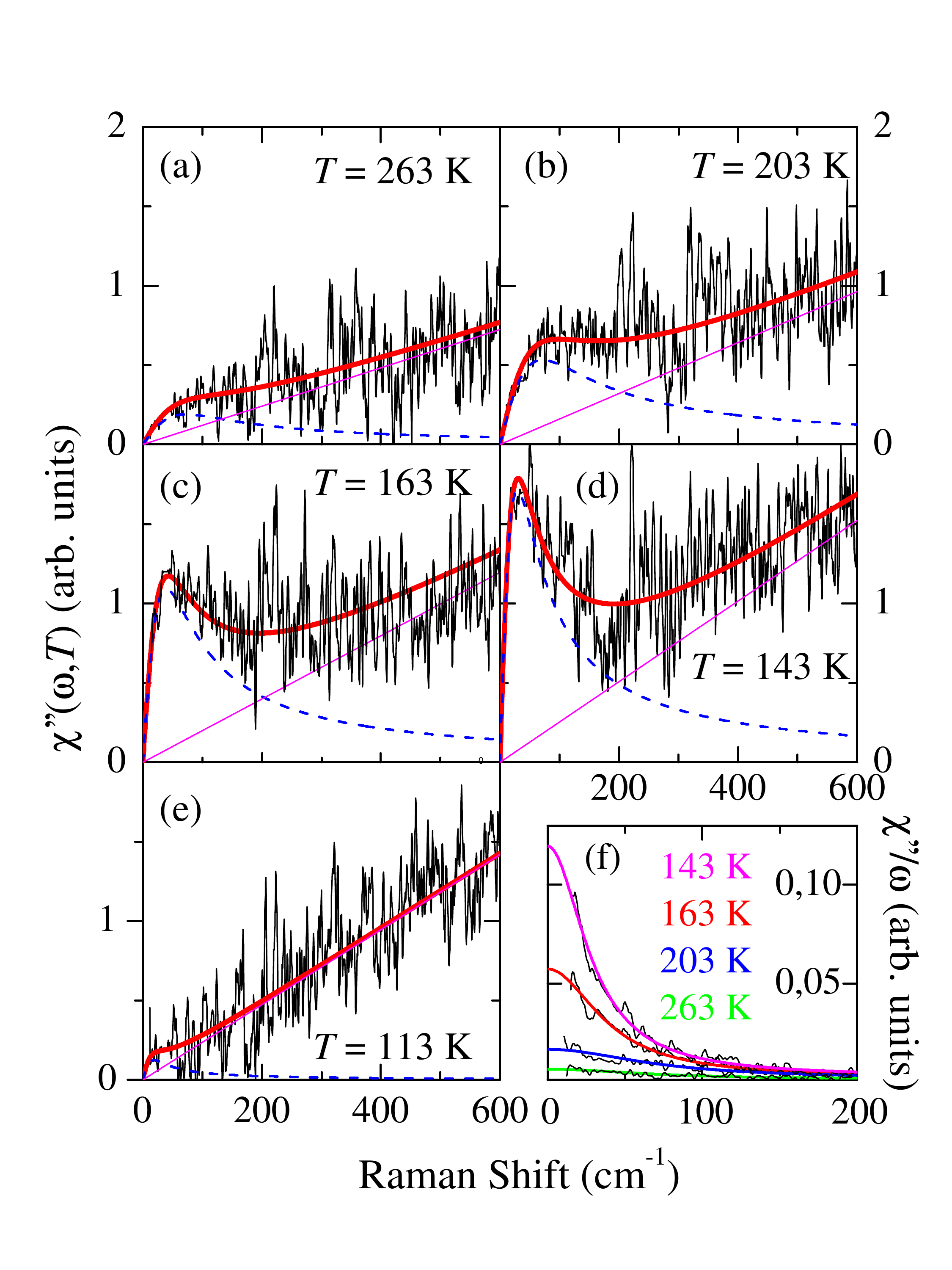}
	\caption{\label{Electronic} (Color online) (a-e)  Raman response $\chi''(\omega,T)$ in $B_{2g}$ symmetry for the tetragonal cell ($XY$ polarization) at selected temperatures. The thick lines are fittings to a model including a Lorentzian quasi-elastic peak (QEP, dashed line) and an additional linear contribution (thin solid line) (see text). (f) $\chi''(\omega,T)/ \omega$ at selected temperatures and corresponding fits to the QEP model.}
\end{figure}

\begin{figure}
	\includegraphics[width=0.5 \textwidth]{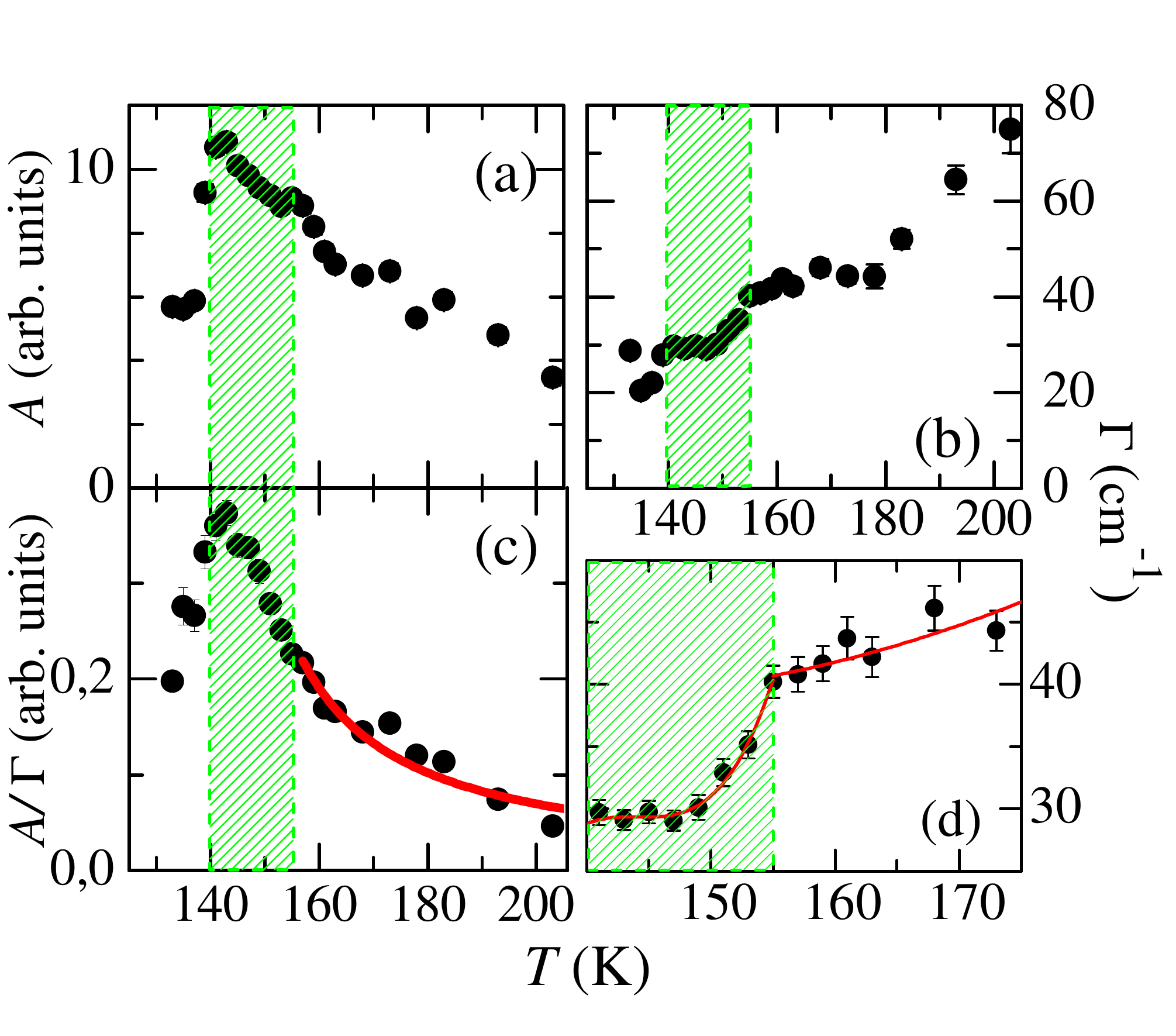}
	\caption{\label{Fluct} (Color online) $T$-dependence of (a) area $A$, (b,d) width $\Gamma$ and (c) height $A/ \Gamma$ of $B_{2g}$ Lorentzian QEP (2 Fe tetragonal cell). Error bars, when not displayed, are smaller than the symbol sizes. The shaded areas mark the $T_{AFM} < T < T_S$ interval. The solid line in (c) shows a fit of the $B_{2g}$ QEP height to a Curie-Weiss-like behavior between $T_S$ and $\sim 200$ K, yielding $\theta_{CW}=137(3)$ K (see text). The solid line in (d) is a guide to the eyes.}
\end{figure}

The Raman response $\chi^{''}(\omega,T)$ is related to the raw intensity $I$ through the relation $I=(1+n) \chi^{''}(\omega,T)+D$, where $n \equiv 1/(e^{\hbar \omega/k_B T}-1)$ is the Bose-Einstein statistical factor and $D$ is an intensity offset (for details, see SM). Figures \ref{Electronic}(a-e) show $\chi^{''} (\omega, T)$ in $XY$ polarization corresponding to $B_{2g}$ symmetry in the 2 Fe tetragonal cell. These measurements were made with much less laser power ($\sim 3$ mW) than for the data shown in Fig. \ref{Phonons} ($\sim 10$ mW), in order to minimize laser heating effects \cite{Kretzschmar}, and were also taken with $4 \times$ less exposure times due to the large number of investigated temperatures. These limitations resulted in poorer signal-to-noise in the data shown in Figs. \ref{Electronic}(a-e). An 8-point-average smoothing is applied in these data for better visualization of the broad electronic Raman signal. A linear component for $\chi_{B_{2g}}^{''}(\omega,T)$ is observed in the frequency region below 600 cm$^{-1}$, which is enhanced below $T_{AFM}$. Measurements performed on an extended frequency region show this component is part of broad peaks at $\sim 2400-3000$ cm$^{-1}$ (see SM). A similar structure was found in BaFe$_2$As$_2$ and attributed to two-magnon scattering \cite{Sugai}. An additional scattering channel, which is most evident at low frequencies ($\omega \lesssim 150$ cm$^{-1}$), is observed in this symmetry and enhances on cooling down to $\sim 140$ K, fading away on further cooling. This contribution is satisfactorily fitted by a quasi-elastic peak (QEP)  $(\chi^{''})_{QEP}^{B_{2g}}(\omega, T) = A(T) \omega \Gamma(T) /(\omega^2+\Gamma(T)^2)$ (dashed lines in Figs. \ref{Electronic}(a-d)), corresponding to a Lorentzian lineshape for $(\chi^{''})_{QEP}^{B_{2g}}(\omega, T) / \omega$. It can be seen from Fig. \ref{Electronic}(f) that the relatively large noise in the Raman response above $\sim 150$ cm$^{-1}$ have little influence on the determination of the QEP fitting parameters $A(T)$ and $\Gamma(T)$. The Raman response for other symmetries accessible by our experimental setup are given in SM at selected temperatures, also showing contributions from two-magnon scattering. 

Figures \ref{Fluct}(a) and \ref{Fluct}(b) show the $T$-dependence of the Lorentzian $B_{2g}$ QEP area $A$ and width $\Gamma$, respectively. Only data between $\sim 120$ and $200$ K are shown, corresponding to the $T-$interval where this signal is sufficiently strong to warrant reliable Lorentzian fits within our statistics. Figure \ref{Fluct}(c) shows $A/ \Gamma$, corresponding to the QEP height, while Fig. \ref{Fluct}(d) is a zoom in of Fig. \ref{Fluct}(b) near $T_S$. Between 280 and 120 K, the $B_{2g}$ QEP area and height show a maximum at $T_{max}=143$ K, slightly above $T_{AFM}$, nearly vanishing below 120 K. Concerning the widths, the $B_{2g}$ QEP gradually sharpens on cooling down to $T_S$. Below $T_S$, $\Gamma_{QEP}$ further sharpens from $\sim 40$ to $\sim 30$ cm$^{-1}$.

\begin{figure}
	\includegraphics[width=0.4 \textwidth]{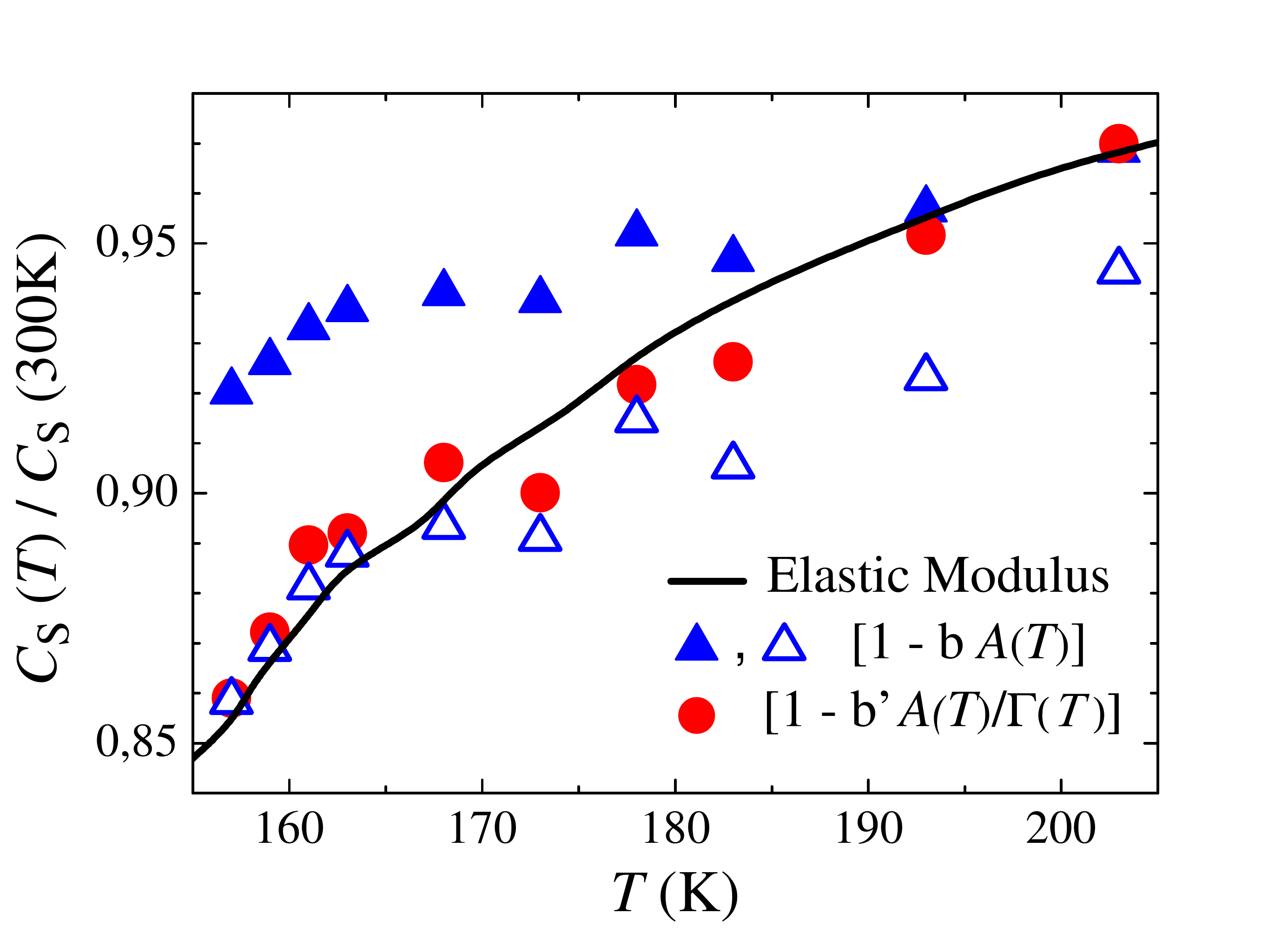}
	\caption{\label{Shear} (Color online) Temperature dependence of the polycrystalline shear modulus taken from Ref. \cite{McGuire} (solid line), and attempted scalings of this curve to the $B_{2g}$ QEP Area $A(T)$ (open and solid triangles) and height $A(T)/ \Gamma(T)$ (circles) extracted from Figs. \ref{Fluct}(a) and \ref{Fluct}(c), respectively.}
\end{figure}

As for the other FeSCs \cite{Gallais,Gallais2,Zhang2016,Massat,Yamase,Kretzschmar,Bohm,Khodas,Yamase2}, we ascribe the $B_{2g}$ QEP in LFAO to electronic nematic fluctuations. In principle, the Raman intensity may be dominated either by charge/orbital or spin nematic fluctuations. The significant residual nematic fluctuations observed between $\sim 120$ K and $T_{AFM}$ (see Figs. \ref{Fluct}(a,c)) are consistent with $^{75}$As NMR measurements that show coexisting AFM and paramagnetic regions in this $T$-interval \cite{Fu}; the paramagnetic regions are expected to host the residual nematic fluctuations observed here. Intriguingly, the temperature where the QEP area and height are maxima, $T_{max}$, does not coincide with the bulk-average $T_S$, contrary to other parent FeSCs \cite{Kretzschmar,Massat}. This deviation is likely related to the broad $T$ interval where tetragonal and orthorhombic domains coexist and fluctuate \cite{Li2}. In this scenario, while the QEP intensity per orthorhombic unit volume is expected to be reduced on cooling, the inverse tendency is found for the remaining tetragonal domains, leading to $T_{max} < T_S$. Still, the nematic fluctuations in LFAO are clearly sensitive to $T_S$, as demonstrated by the sharpening of the $B_{2g}$ QEP below $T_S$ (see Fig. \ref{Fluct}(d)). In fact, this is a manifestation of slower nematic fluctuations in the orthorhombic phase. This is again qualitatively consistent with $^{75}$As NMR results that show a slowing down of the magnetic dynamics below $T_S$ \cite{Fu} and may be also related with the enhancement of the magnetic correlation length below $T_S$ observed in inelastic neutron scattering measurements \cite{Zhang}.

We discuss our results considering separately the independent scenarios where charge/orbital or spin nematic fluctuations dominate the intensity of the $B_{2g}$ Raman QEP. Starting with the charge/orbital scenario (scenario A), the bare static nematic susceptibility $\chi_{nem}^{(0)}(T)$ and $(\chi^{''})^{B_{2g}}_{QEP}(\omega,T)$ are directly connected by a Kramers-Kronig transformation $\chi_{nem}^{(0)}(T) = (2/ \pi) \int_0 ^{ \infty} (\chi^{''})^{B_{2g}}_{QEP}(\omega,T) / \omega d \omega$ \cite{Gallais,Gallais2}, corresponding to the QEP area $A(T)$ in our analysis. An attempted scaling of $\chi_{nem}^{(0)}(T)$ obtained in this way and the polycrystalline shear modulus $C_S$ extracted from Ref.\cite{McGuire}, i.e., $C_{S}(T)/C_{S}$(300 K)$ = 1 - b A(T)$ \cite{Gallais2}, is given in Fig. \ref{Shear}, where $b$ is a free parameter (see footnote \cite{explanation}). In our analysis, we tentatively varied $b$ to scale $A(T)$ to $C_{S}(T)$ either at $T \gtrsim T_S$ (empty triangles in Fig. \ref{Shear}) or at $T \sim 200$ K $>> T_S$ (filled triangles). However, no value for $b$ yielded a satisfactory scaling for the entire investigated interval $T_S < T \lesssim 200$ K. The lack of scaling between the shear modulus and the QEP area, interpreted under scenario A, indicate that the charge/orbital fluctuations do not drive the structural transition at $T_S$, and an additional electronic nematic degree of freedom, presumably the magnetic one, is driving the phase transitions in LFAO \cite{Fernandes}. This reasoning closely follows that presented in Ref. \cite{Gallais} for BaFe$_2$As$_2$.

We now explore the alternative scenario where spin nematic fluctuations dominate the intensity of the $B_{2g}$ Raman QEP (scenario B). In this case, the dynamical electronic nematic susceptibility is not given directly by $(\chi^{''})^{B_{2g}}_{QEP}(\omega,T)$, and therefore a Kramers-Kronig transformation does not apply to extract $\chi_{nem}^{(0)}(T)$. Instead, $\chi_{nem}^{(0)}(T)$ is proportional to the slope of $(\chi^{''})^{B_{2g}}_{QEP}(\omega,T)$ in the limit $\omega \rightarrow 0$ \cite{Kretzschmar,Karahasanovic}, namely the QEP height $A(T)/ \Gamma(T)$. In this scenario, $\Theta_{in}=137(3)$ K, obtained from the fit of $A(T)/ \Gamma(T)$ to a Curie-Weiss-like behavior $A/ \Gamma = C/(T-\Theta_{in})$ over the interval $T_S < T \lesssim 200 $ K (solid line in Fig. \ref{Fluct}(c)), is the bare nematic transition temperature in the absence of the magneto-elastic coupling that induces the transition at higher temperatures. Figure \ref{Shear} displays a scaling of the polycrystalline shear modulus to the peak height, $C_{S}(T)/C_{S}$(300 K)$ = 1 - b' A(T)/\Gamma(T)$, showing an excellent agreement for the entire investigated interval. Therefore, independently of the assumption on the detailed nature of the Raman $B_{2g}$ QEP, our analysis supports the scenario where the nematic transition is magnetically driven.

The thermal evolution of the relaxation rate $\Gamma^{B_{2g}}$ provides further insight into the nematic transition. At $T \sim 200$ K one has $\Gamma^{B_{2g}} \sim 10$ meV ($\sim$ 80 cm$^{-1}$), see Fig. \ref{Fluct}(b), which is on the same energy scale of the optical phonons (see Fig. \ref{Phonons}). However, the nematic fluctuations slow down continuously on cooling (see Fig. \ref{Fluct}(b)). Presumably, as the nematic fluctuation rate become significantly smaller than the typical optical phonon frequencies, local and instantaneous orthorhombic distortions are expected to rise and accompany the electronic nematic correlations. We suggest that at $T_S$ the growing lattice strain caused by the local orthorhombic distortions finally drive the formation of a long-range orthorhombic phase, i.e., the so-called nematic phase. Immediately below $T_S$ the nematic fluctuations are slowed down further (see Fig. \ref{Fluct}(d)). This is likely associated with changes in the $J_a$ and $J_b$ nearest-neighbor exchange integrals, partially releasing the magnetic frustration and allowing for increased magnetic correlation lengths \cite{Zhang}.

Further inspection of our results gives insight into the large separation between $T_S$ and $T_{AFM}$ (~15 K) compared to their near coincidence in BaFe$_2$As$_2$. We note that at $T=163$ K, for instance, the maximum of $\chi^{''}_{B_{2g}}(\omega,T)$, corresponding to the QEP linewidth $\Gamma$, is 43(2) cm$^{-1}$ for LFAO (see Fig. \ref{Electronic}(c) and \ref{Fluct}(d)), much smaller than $\sim 100$ cm$^{-1}$ for BFA at this temperature \cite{Gallais}. Such slower nematic fluctuations in LFAO preempt the stabilization of orthorhombic domains significantly above $T_{AFM}$. 
This scenario may also give insight into the nematic transition of other FeSCs. For instance, a $B_{2g}$ QEP has also been reported \cite{Massat} for FeSe, which also gradually sharpens on cooling, reaching $\Gamma \sim 30$ cm$^{-1}$ at $T_S$, which is comparable to the observed $\Gamma$ for LFAO in the nematic phase (see Fig. \ref{Fluct}(d)).

In summary, polarized Raman scattering in LaFeAsO reveals a quasi-elastic $B_{2g}$ scattering channel from nematic fluctuations above $\sim T_{AFM}$. An analysis of the $T$-dependence of this signal supports the conclusion that magnetism is the primary order parameter driving the phase transitions in this material. Relatively slow electronic nematic fluctuations preempt $T_S$ and arguably signal the separation between $T_S$ and $T_{AFM}$.

We thank R. M. Fernandes and P. G. Pagliuso for their critical reading of this manuscript and helpful discussions. This work was supported by FAPESP Grant 2012/04870-7 and CNPq, Brazil. Ames Laboratory is supported by the US Department of Energy, Office of Basic Energy Sciences, Division of Materials Sciences and Engineering under Contract No. DE-AC02-07CH11358.

\newpage

\begin{appendix}

\section{Supplemental Material for ``Nematic Fluctuations and Phase Transitions in LaFeAsO: a Raman Scattering Study'' by U. F. Kaneko {\it et al.}}

\subsection{Temperature-dependence of the Arsenic $A_g$ phonon}

\renewcommand{\thefigure}{S\arabic{figure}}

\setcounter{figure}{0}

\begin{figure}
	\includegraphics[width=0.5 \textwidth]{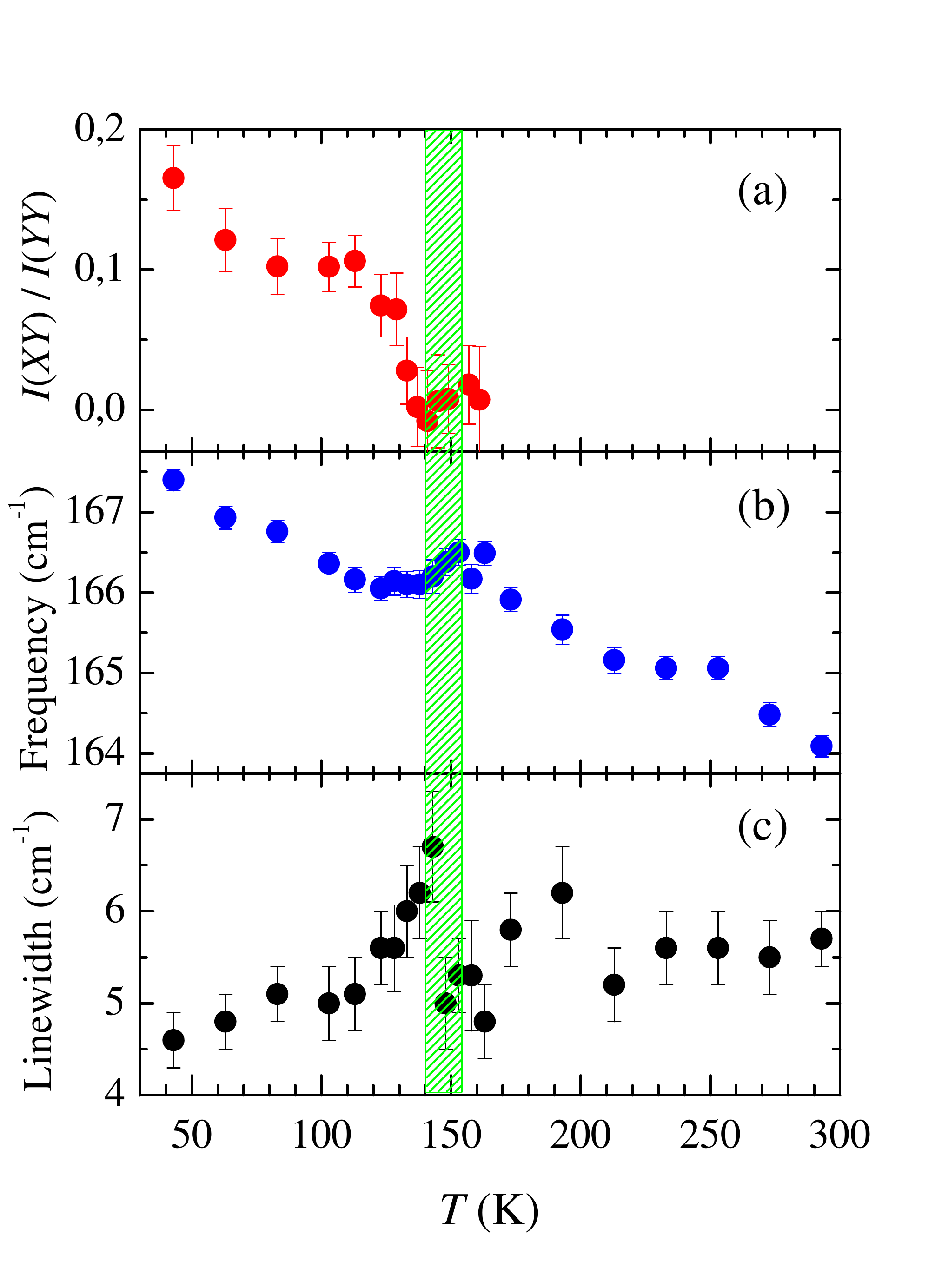}
	\caption{Temperature dependence of the As $A_g$ mode: (a) intensity in $XY$ polarization relative to $YY$ polarization ($I_{XY}/I_{YY}$), (b) frequency, and (c) linewidth. The shaded area marks the $T_{AFM} < T < T_S$ interval.} \label{Tdep}
\end{figure}

Figure \ref{Tdep}(a) shows the temperature-dependence of the relative intensity of the As mode in $XY$ with respect to $YY$ polarizations (see also Fig. 1 of the main text). An enhancement of this mode in $XY$ polarization is observed below $T_{AFM}$. This is similar to observed in related systems \cite{Chauviere3,Choi,Chauviere2,Kretzschmar} and can be ascribed to the coupling of this phonon with anisotropic electronic states in the magnetic phase. \cite{Martinez} Figures \ref{Tdep}(b) and \ref{Tdep}(c) show the frequency $\omega_0^{As}$ and linewidth $\Gamma^{As}$ of this mode. Frequency anomalies are observed at both $T_S$ and $T_{AFM}$, the latter being likely due to spin-phonon coupling, \cite{Granado} while the phonon linewidth shows a maximum at $T \sim T_{AFM}$ with no anomaly at $T_S$. The linewidth of this mode at low-$T$ is resolution-limited, suggesting a high crystalline quality an homogeneity, at least within the relatively small sample region probed by the laser spot.

\subsection{Extraction of the Raman response from the raw intensities}

\begin{figure}
	\includegraphics[width=0.5 \textwidth]{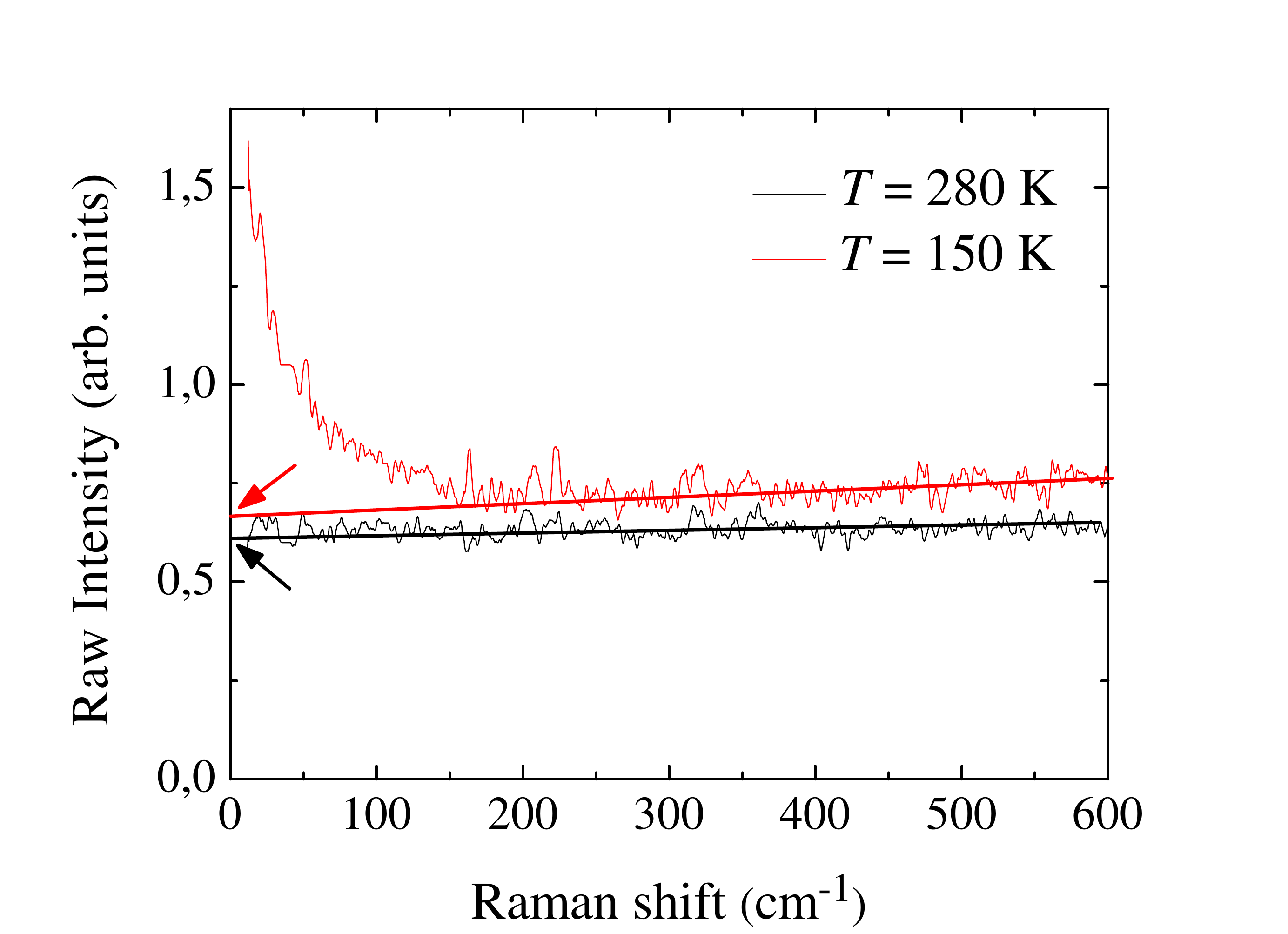}
	\caption{\label{raw} Raw intensity spectra in $XY$ polarization at $T=280$ and 150 K. The solid lines show linear fits to the data above 300 cm$^{-1}$ and the arrows mark the extrapolation of these lines at $\omega \rightarrow 0$, taken as the intensity offset $D(T)$.}
\end{figure}

As discussed in the main text, the raw intensity $I$ (after subtraction of the dark noise) and the Raman response  $\chi^{''}(\omega,T)$ are related by the expression $I=(1+n) \chi^{''}(\omega,T)+D$, where $n$ is the Bose-Einstein statistical factor and $D$ represents an intensity offset. The offset, which is frequency-independent over the limited spectral interval of interest to this work ($\omega < 600 $ cm$^{-1}$), may be due to a combination of residual stray light in the spectrograph stage of our instrument and luminescence.  $D$ was subtracted from the raw intensity prior to the Bose-Einstein correction in order to avoid distorting the Raman susceptibility spectra. Non-systematical variations of $D$ the order of 20 \% were observed as the laser spot moved slightly along the sample surface at different temperatures. Figure \ref{raw} shows the raw intensity spectra in $XY$ polarization for $T=280$ K and 150 K. At $T=280$ K, the intensity follows a linear $\omega-$dependence down to the lowest frequencies. This indicates that no significant intrinsic Raman intensity is present for $\omega \rightarrow 0$ at $T=280$ K, which otherwise would be highly amplified by the Bose-Einstein statistical factor in the low-$\omega$ region leading to a non-linear contribution to the raw intensity. Thus, the raw intensity for $\omega \rightarrow 0$ at $T=280$ K was taken as the offset $D(280$ K). At $T=150$ K, a significant quasi-elastic intensity is also observed, associated with the nematic fluctuations (see main text). The offset $D(150$ K) was then taken as the extrapolation to $\omega =0$ of a linear fit taken in the region $300 < \omega < 600$ cm$^{-1}$ (see Fig. \ref{raw}). This procedure was repeated for all investigated temperatures and polarizations.

\subsection{Raman response at different polarizations; two-magnon scattering}

\begin{figure}
	\includegraphics[width=0.5 \textwidth]{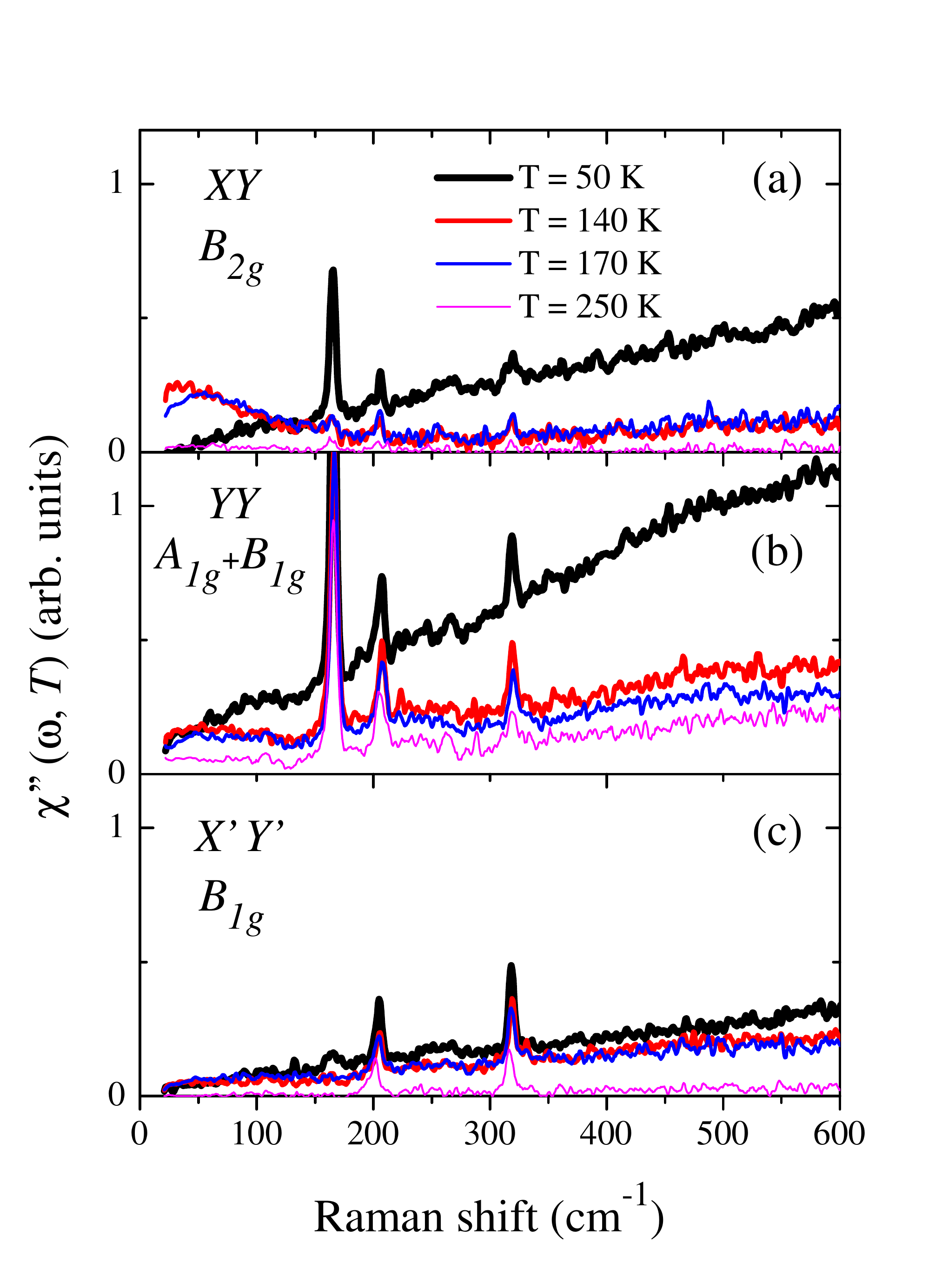}
	\caption{\label{pol} Raman response $\chi''(\omega,T)$ at selected temperatures for $XY$ (a), $YY$ (b) and $X'Y'$ (c) polarizations. The corresponding symmetries defined according to the tetragonal unit cell are indicated.}
\end{figure}

\begin{figure}
	\includegraphics[width=0.5 \textwidth]{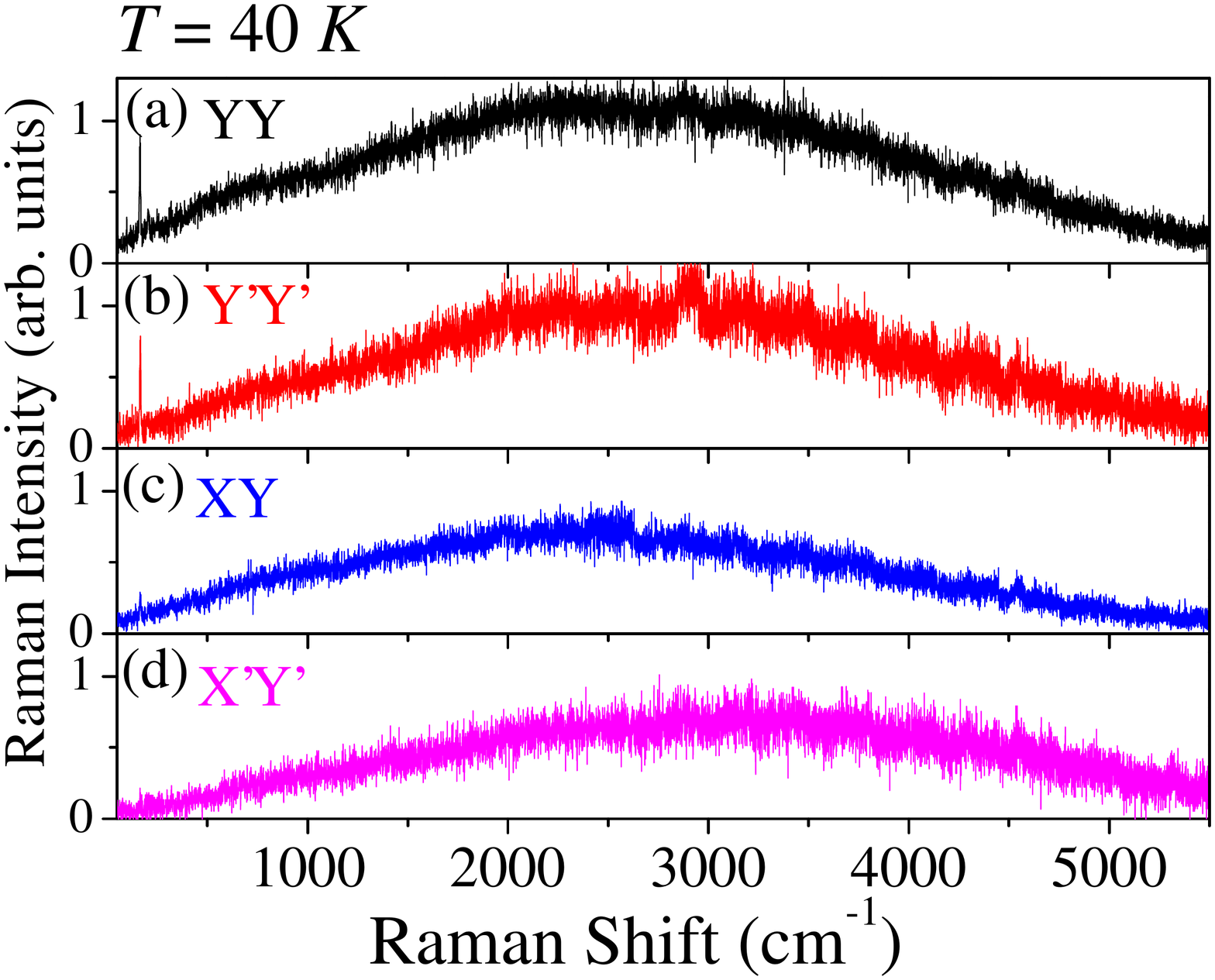}
	\caption{\label{magnons1} Polarized Raman spectra over an extended spectral interval at $T=40$ K.}
\end{figure}

\begin{figure}
	\includegraphics[width=0.5 \textwidth]{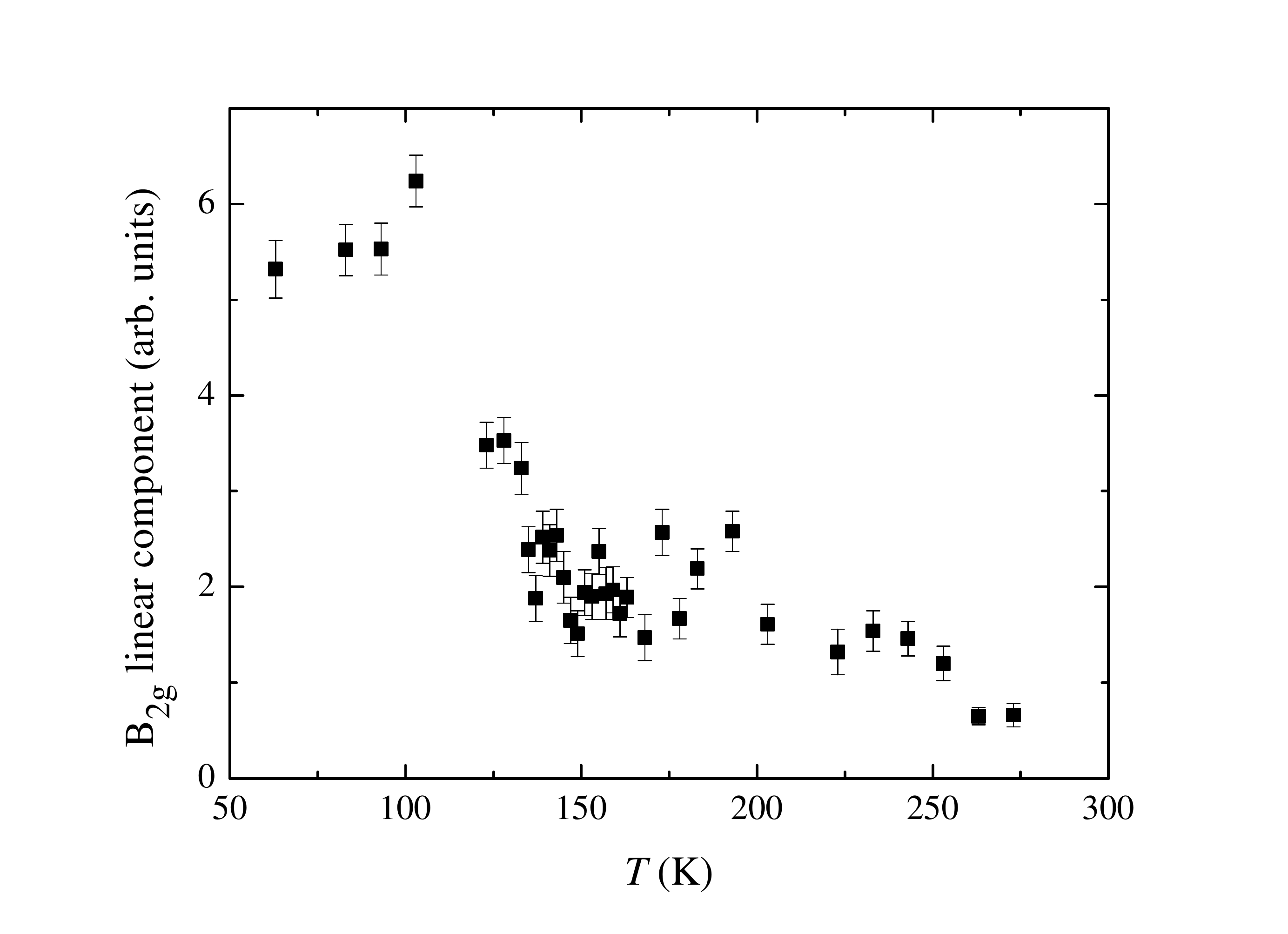}
	\caption{\label{linear} Temperature-dependence of the slope of the linear component of $\chi''(\omega,T)$ extracted in the frequency region below 600 cm$^{-1}$ for $XY$ polarization (see also Fig. 2 of the main text and Figs. \ref{magnons1} and \ref{pol}).}
\end{figure}

Although the focus of this work is on the $B_{2g}$ quasi-elastic peak (QEP) obtained in $XY$ polarization, the electronic Raman signal of LaFeAsO was also investigated for the other accessible linear polarizations. Figures \ref{pol}(a-c) show the Raman response for $B_{2g}$, $A_{1g}+B_{1g}$ and $B_{1g}$ symmetries, respectively, at selected temperatures and $\omega < 600$ cm$^{-1}$. Besides the phonon modes and the $B_{2g}$ QEP discussed in the main text, an additional signal that increases almost linearly with frequency in this range is observed for all polarizations. Figures \ref{magnons1}(a-d) show the spectra at $T=40$ K over an extended frequency region. Broad peaks were observed at $\sim 2400$ and $\sim 3100$ cm$^{-1}$ for $XY$ and $X'Y'$ polarizations, respectively, while the spectra at $YY$ and $Y'Y'$ polarizations seem to contain a combination of these contributions. It is important to notice that the linear component observed for $\omega < 600$ cm$^{-1}$ in $XY$ polarization is actually the lower-frequency limit of this broad scattering.  Further insight into the nature of this contribution is gained by the $T$-dependence of the slope of the linear component in the frequency range 300 $< \omega < 600$ cm$^{-1}$ in $XY$ polarization, using the data of Fig. 2 of the main text. Notably, this slope increases significantly below $T_N$. Thus, the linear contribution is not associated with nematic fluctuations, and can be safely excluded from the computation of the instantaneous nematic susceptibility. In fact, for BaFe$_2$As$_2$, broad signals centered at high frequencies were also observed and ascribed to two-magnon scattering. \cite{Sugai} This identification is consistent with the temperature-dependence shown in Fig. \ref{linear}.

\end{appendix}


\begin{references}
	
	\bibitem{Ge} J.-F. Ge, Z.-L. Liu, C. Liu, C.-L. Gao, D. Qian, Q.-K. Xue, Y. Liu and J.-F. Jia, Nature Mater. {\bf 14}, 285 (2015).
	
	\bibitem{Kamihara} Kamihara, T. Watanabe, M. Hirano, and H. Hosono, J. Am. Chem. Soc. {\bf 130}, 3296 (2008).
	
	\bibitem{Paglione} J. Paglione and R.L. Greene, Nature Phys. {\bf 6} 645 (2010).
	
	\bibitem{Fernandesnem} R.M. Fernandes and J. Schmalian, Supercond. Sci. Technol. {\bf 25}, 084005 (2012).
	
	\bibitem{Fernandes} R.M. Fernandes, A.V. Chubukov, and J. Schmalian, Nature Phys. {\bf 10}, 97 (2014).
	
	
	\bibitem{Chu} J.-H. Chu, J.G. Analytis, K. De Greeve, P.L. McMahon, Z. Islam, Y. Yamamoto, and I.R. Fisher, Science {\bf 329}, 824 (2010).
	
	\bibitem{Ando} Y. Ando, K. Segawa, S. Komiya, and A.N. Lavrov, Phys. Rev. Lett. {\bf 88}, 137005 (2002).
	
	\bibitem{Kuo} H.-H. Kuo, J.-H. Chu, J.C. Palmstrom, S.A. Kivelson, and I.R. Fisher, Science {\bf 352}, 958 (2016).
	
	\bibitem{Kim} M.G. Kim, R.M. Fernandes, A. Kreyssig, J.W. kim, A. Thaler, S.L. Bud'ko, P.C. Canfield, R.J. McQueeney, J. Schmalian, and A.I. Goldman, Phys. Rev. B {\bf 83}, 134522 (2011).
	
	\bibitem{Garitezi} T.M. Garitezi, C. Adriano, P.F.S. Rosa, E.M. Bittar, L. Bufaical, R.L. de Almeida, E. Granado, T. Grant, Z. Fisk, M.A. Avila, R.A. Ribeiro, P.L. Kuhns, A.P. Reyes, R.R. Urbano, and P.G. Pagliuso, Braz. J. Phys. {\bf 43}, 223 (2013).
	
	\bibitem{McGuire} M.A. McGuire, A.D. Christianson, A.S. Sefat, B.C. Sales, M.D. Lumsden, R. Jin E.A. Payzant, D. Mandrus, Y. Luan, V. Keppens, V. Varadarajan, J.W. Brill, R.P. Hermann, M.T. Sougrati, F. Grandjean, and G.J. Long, Phys. Rev. B {\bf 78}, 094517 (2008).
	
	\bibitem{Yan2009} J.-Q. Yan, S. Nandi, J.L. Zarestky, W. Tian, A. Kreyssig, B. Jensen, A. Kracher, K.W. Dennis, R.J. McQueeney, A.I. Goldman, R.W. McCallum, and T.A. Lograsso, Appl. Phys. Lett. {\bf 95}, 222504 (2009).
	
	\bibitem{Li} H.-F. Li, W. Tian, Q.-Q. Yan, J.L. Zarestky, R.W. McCallum, T.A. Lograsso, and D. Vaknin, Phys. Rev. B {\bf 82}, 064409 (2010).
	
	
	\bibitem{McQueen} T.M. McQueen, A.J. Williams, P.W. Stephens, J. Tao, Y. Zhu, V. Ksenofontov, F. Casper, C. Felser, and R.J. Cava, Phys. Rev. Lett. {\bf 103}, 057002 (2009).
	
	\bibitem{Pratt} D.K. Pratt, W. Tian, A. Dreyssig, J.L. Zarestky, S. Nandi, N. Ni, S.L. Bud'ko, P.C. Canfield, A.I. Goldman, and R.J. McQueeney, Phys. Rev. Lett. {\bf 103}, 087001 (2009).
	
	\bibitem{Baek} S.-H. Baek, D.V. Efremov, J.M. Ok, J.S. Kim, J. van den Brink, and B. B\"uchner, Nature Mater. {\bf 14}, 210 (2014).
	
	\bibitem{Bohmer} A.E. B\"ohmer, T. Arai, F. Hardy, T. Hattori, T. Iye, T. Wolf, H. v. L\"ohneysen, K. Ishida, and C. Meingast, Phys. Rev. Lett. {\bf 114}, 027001 (2015).
	
	\bibitem{Chubukov} A.V. Chubukov, R.M. Fernandes, and J. Schmalian, Phys. Rev. B {\bf 91}, 201105(R) (2015). 
	
	\bibitem{Chauviere} L. Chauvi\`ere, Y. Gallais, M. Cazayous, M.A. M\'easson, A. Sacuto, D. Colson, and A. Forget Phys. Rev. B {\bf 82}, 180521(R) (2010).
	
	\bibitem{Chauviere3} L. Chauvi\`ere, Y. Gallais, M. Cazayous, M.A. M\'easson, A. Sacuto, D. Colson, and A. Forget, Phys. Rev. B {\bf 84}, 104508 (2011).
	
	\bibitem{Gallais} Y. Gallais, R.M. Fernandes, I. Paul, L. Chauvi\`ere, Y.-X. Yang, M.-A. M\'easson, M. Cazayous, A. Sacuto, D. Colson, and A. Forget, Phys. Rev. Lett. {\bf 111}, 267001 (2013).
	
	\bibitem{Gallais2} Y. Gallais and I. Paul, C. R. Physique {\bf 17}, 113 (2016).
	
	\bibitem{Kretzschmar} F. Kretzschmar, T. B\"ohm, U. Karahasanovi\'c, B. Muschler, A. Baum, D. Jost, J. Schmalian, S. Caprara, M. Grilli, C. Di Castro, J.G. Analytis, J.-H. Chu, I.R. Fisher, and R. Hackl, Nature Phys. {\bf 12}, 560 (2016).
	
	\bibitem{Zhang2016} W.-L. Zhang, Z.P. Yin, A. Ignatov, Z. Bukowski, J. Karpinski, A.S. Sefat, H. Ding, P. Richard, and G. Blumberg, Phys. Rev. B {\bf 93}, 205106 (2016).
	
	\bibitem{Bohm} T.B\"ohm, R. Hosseinian Ahangharnejhad, D. Jost, A. Baum, B. Muschler, F. Kretzschmar, P. Adelmann, T. Wolf, H.-H. Wen, J.-H. Chu, I.R. Fisher, and R. Hackl, arXiv: 1608.02772 (2016).
	
	\bibitem{Massat} P. Massat, D. Farina, I. Paul, S. Karlsson, P. Strobel, P. Toulemonde, M.-A. M\'easson, M. Cazayous, A. Sacuto, S. Kasahara, T. Shibauchi, Y. Matsuda, and Y. Gallais, PNAS {\bf 113}, 9177 (2016).
	
	\bibitem{Gnezdilov} V. Gnezdilov, Y.G. Pashkevich, P. Lemmens, D. Wulferding, T. Shevtsova, A. Gusev, D. Chareev, and A. Vasiliev, Phys. Rev. B {\bf 87}, 144508 (2013).
	
	\bibitem{Thorsmolle} V.K. Thorsmolle, M. Khodas, Z.P. Yin, C. Zhang, S.V. Carr, P. Dai, and G. Blumberg, Phys. Rev. B {\bf 93}, 054515 (2016).
	
	\bibitem{Yamase} H. Yamase and R. Zeyler, Phys. Rev. B {\bf 88}, 125120 (2013).
	
	\bibitem{Khodas} M. Khodas and A. Levchenko, Phys. Rev. B {\bf 91}, 235119 (2015).
	
	\bibitem{Yamase2} H. Yamase and R. Zeyher, New J. Phys. {\bf 17}, 073030 (2015).
	
	\bibitem{Zhang} Q. Zhang, R.M. Fernandes, J. Lamsal, J. Yan, S. Chi, G.S. Tucker, D.K. Pratt, J.W. Lynn, R.W. McCallum, P.C. Canfield, T.A. Lograsso, A.I. Goldman, D. Vaknin, and R.J. McQueeney, Phys. Rev. Lett. {\bf 114}, 057001 (2015).
	
	\bibitem{Hadjiev} V.G. Hadjiev, M.N. Iliev, K. Sasmal, Y.-Y. Sun, and C.W. Chu, Phys. Rev. B {\bf 77}, 220505(R) (2008).
	
	\bibitem{Um} Y.J. Um, Y. Bang, B.H. Min, Y.S. Kwon, and M. Le Tacon, Phys. Rev. B {\bf 89}, 184510 (2014).
	
	\bibitem{Choi} K.-Y. Choi, D. Wulferding, P. Lemmens, N. Ni, S.L. Bud'ko, and P.C. Canfield, Phys. Rev. B {\bf 78}, 212503 (2008).
	
	\bibitem{Litvinchuk} A.P. Litvinchuk, V.G. Hadjiev, M.N. Iliev, B. Lv, A.M. Guloy, and C.W. Chu, Phys. Rev. B {\bf 78}, 060503(R) (2008). 
	
	\bibitem{Chauviere2} L. Chauvi\'ere, Y. Gallais, M. Cazayous, A. Sacuto, M.A. M\'easson, D. Colson, and A. Forget, Phys. Rev. B {\bf 80}, 094504 (2009).
	
	\bibitem{Martinez} N.A. Garc\'ia-Mart\'inez, B. Valenzuela, S. Ciuchi, E. Cappelluti, M.J. Calder\'on, and E. Bascones, Phys. Rev. B {\bf 88}, 165106 (2013).
	
	\bibitem{Sugai} S. Sugai, Y. Mizuno, R. Watanabe, T. Kawaguchi, K. Takenaka, H. Ikuta, Y. Takayanagi, N. Hayamizu, and Y. Sone, J. Phys. Soc. Jpn. {\bf 81}, 024718 (2012).
	
	\bibitem{Fu} M. Fu, D.A. Torchetti, T. Imai, F.L. Ning, J.-Q. Yan, and A.S. Sefat, Phys. Rev. Lett. {\bf 109}, 247001 (2012).
	
	\bibitem{Li2} H.-F. Li, J.-Q. Yan, J.W. Kim, R.W. McCallum, T.A. Lograsso, and D. Vaknin, Phys. Rev. B {\bf 84}, 220501(R) (2011).
	
	\bibitem{explanation} There is no report on single crystal shear moduli in LFAO, to the best of our knowledge. The polycrystalline shear modulus ``$C_{44}$'' shown in ref. \cite{McGuire} is a superposition of the several single crystal shear moduli. However, only $C_{66}$ is expected to become critical and display strong temperature dependence, and in our analysis we assume that it dominates the temperature-dependence of the polycrystalline shear modulus.
	
	\bibitem{Karahasanovic} U. Karahasanovic, F. Kretzschmar, T. B\"ohm, R. Hackl, I. Paul, Y. Gallais, and J. Schmalian, Phys. Rev. B {\bf 92}, 075134 (2015).
	
	\bibitem{Fernandes2} R.M. Fernandes, L.H. VanBebber, S. Bhattacharya, P. Chandra, V. Keppens, D. Mandrus, M.A. McGuire, B.C. Sales, A.S. Sefat, and J. Schmalian, Phys. Rev. Lett. {\bf 105}, 157003 (2010).

	
	
\end{references}

\begin{references}
	
	\bibitem{Chauviere3} L. Chauvi\`ere, Y. Gallais, M. Cazayous, M.A. M\'easson, A. Sacuto, D. Colson, and A. Forget, Phys. Rev. B {\bf 84}, 104508 (2011).
	
	\bibitem{Choi} K.-Y. Choi, D. Wulferding, P. Lemmens, N. Ni, S.L. Bud'ko, and P.C. Canfield, Phys. Rev. B {\bf 78}, 212503 (2008).
	
	\bibitem{Chauviere2} L. Chauvi\'ere, Y. Gallais, M. Cazayous, A. Sacuto, M.A. M\'easson, D. Colson, and A. Forget, Phys. Rev. B {\bf 80}, 094504 (2009).
	
	\bibitem{Kretzschmar} F. Kretzschmar, T. B\"ohm, U. Karahasanovi\'c, B. Muschler, A. Baum, D. Jost, J. Schmalian, S. Caprara, M. Grilli, C. Di Castro, J.G. Analytis, J.-H. Chu, I.R. Fisher, and R. Hackl, Nature Phys. {\bf 12}, 560 (2016).
	
	\bibitem{Martinez} N.A. Garc\'ia-Mart\'inez, B. Valenzuela, S. Ciuchi, E. Cappelluti, M.J. Calder\'on, and E. Bascones, Phys. Rev. B {\bf 88}, 165106 (2013).
	
	\bibitem{Granado} E. Granado, A. Garc\'ia, J.A. Sanjurjo, C. Rettori, I. Torriani, F. Prado, R.D. S\'anchez, A. Caneiro, and S.B. Oseroff, Phys. Rev. B {\bf 60}, 11879 (1999).
	
	
	\bibitem{Sugai} S. Sugai, Y. Mizuno, R. Watanabe, T. Kawaguchi, K. Takenaka, H. Ikuta, Y. Takayanagi, N. Hayamizu, and Y. Sone, J. Phys. Soc. Jpn. {\bf 81}, 024718 (2012).
	
\end{references}
\end{document}